\documentclass[preprint2]{aastex}

\shorttitle{The Interacting Early-Type Binary V382 Cyg}
\shortauthors{B . Ya{\c s}arsoy, and K. Yakut}

\begin{document}

\title{The Interacting Early-Type Binary V382 Cyg}
\author{B. Ya{\c s}arsoy\altaffilmark{1} and K. Yakut\altaffilmark{1,2}}
\affil{$^1$Department of Astronomy and Space Sciences, University of Ege, 35100, Bornova--{\.I}zmir, Turkey}
\affil{$^2$Institute of Astronomy, University of Cambridge, Madingley Road, Cambridge CB3 0HA}

\begin{abstract}
We present photometric and spectroscopic data analysis and orbital period study
of an early-type interacting binary system V382~Cyg by using all the available data.
We made a simultaneous light and radial velocity curve solution. The derived physical parameters of the primary and secondary stellar components are
$M_{1}$ = 27.9(5) $M_{\odot}$, $M_{2}$ = 20.8(4) $M_{\odot}$, $R_{1}$ = 9.7(2) $R_{\odot}$,
$R_{2}$ = 8.5(2) $R_{\odot}$, $\log{(L_1/L_{\odot})} = 5.152(20)$ and $\log{(L_2/L_{\odot})}  = 4.954(19)$
while the separation of the components is {\it a} = 23.4 $R_{\odot}$.
Newly obtained parameters yield the distance of the system as 1466(76) pc.
Analyses of the mid-eclipse times indicate a period increase of  $\frac{dP}{dt}=4.2(1)\times 10^{-7}$ days/yr
that can be interpreted in terms of the high mass transfer ($\frac{dM}{dt}=6.1(5)\times 10^{-6}$ $M_{\odot}$/yr)
from the less massive  component to the more massive component.
Finally we modelled the evolution of the components using nonconservative codes and discussed the obtained results.
The age of the binary system is estimated  as 3.85 Myr.
\end{abstract}


\section{Introduction}

Massive interacting binary systems are the progenitor candidates of some high energy phenomena in astrophysics (e.g., supernovae).
Solar-type close binary systems with convective envelopes can evolve in different ways than close massive stars.
Yakut \& Eggleton (2005) modelled late-type close binary systems assuming conservative and non-conservative cases.
The authors proposed that a detached configuration can evolve into a semi-detached configuration followed by a contact configuration.
Except late-type systems, we have less information on the evolution of the interactive massive binary systems,
since there are a few binary systems with well-determined parameters.
Therefore, systems like V382 Cyg, RY Sct (Djurasevic et al. 2008) and V729 Cyg (Yasarsoy et al. 2013, submitted),  etc. 
are crucial in observational studies of close massive binaries.

Photometric studies of V382 Cyg have been published by different researchers.
The binary system V382 Cyg was discovered as a variable star by Morgentoth (1935) and Petrov (1946) and they classified the system as an eclipsing binary.
The photometric UBV light variations of the system was obtained by Landolt (1964) by using 16-inch KPNO telescope.
Modelling the light and radial velocity curves of the system, Bloomer et al. (1979) derived an orbital inclination of 87$^{o}$ and
$M_{1}$ = 26.7 (5)$M_{\odot}$ and $M_{2}$ = 18.9 (4)$M_{\odot}$.
Later photometric studies were performed by Cester et al. (1978), Mayer et al. (1986), Harries et al. (1997), De{\v g}irmenci et al. (1999) and Qian et al. (2007).

Spectroscopic studies of the system were provided by Pearce (1952), Popper (1978), Koch et al. (1979), Harries et al. (1997), Burkholder et al. (1997) and Mayer et al. (2002).
Popper (1978) studying the He I and He II lines derived the semi-amplitude of the radial velocity curve and the mass function for the components
as $K_1=255$ km\,s$^{-1}$, $K_2=360$ km\,s$^{-1}$, $M_1\sin^3 i = 26.7$ $M_{\odot}$ and $M_2\sin^3 i = 18.9$ $M_{\odot}$.
Later on Popper and Hill (1991) recalculated the parameters as $K_1=276$ km\,s$^{-1}$ and $K_2=400$ km\,s$^{-1}$ and the masses as $M_{1}$ = 32.6 (1.8)$M_{\odot}$ and  $M_{2}$ = 22.9(1.3)$M_{\odot}$.
By using photometric and spectroscopic data Harries et al. (1997) derived the mass of the components as
$M_{1}$ = 26.0 (7)$M_{\odot}$ and  $M_{2}$ = 19.3 (4)$M_{\odot}$.
Recently, Mayer et al. (2002) derived the masses for the stars as $M_{1}$ = 29.2 $M_{\odot}$ and  $M_{2}$ = 21.2 $M_{\odot}$.
Spectral types of the components were defined by Pearce (1952) as O6.5+O7.5, by Popper (1980) as O7.3+O7.7, by Harries et al. (1997) as O7.3+O7.7 and by Burkholder et al. (1997) as O6.5V(f) + O6V(f). They gave distance module for the system as 11.5(5) mag.
Garmany \& Stencel (1992) classified V382 Cyg as a member of Cyg OB1 association.

\section{New Observations}

\begin{figure}
\includegraphics[width=80mm]{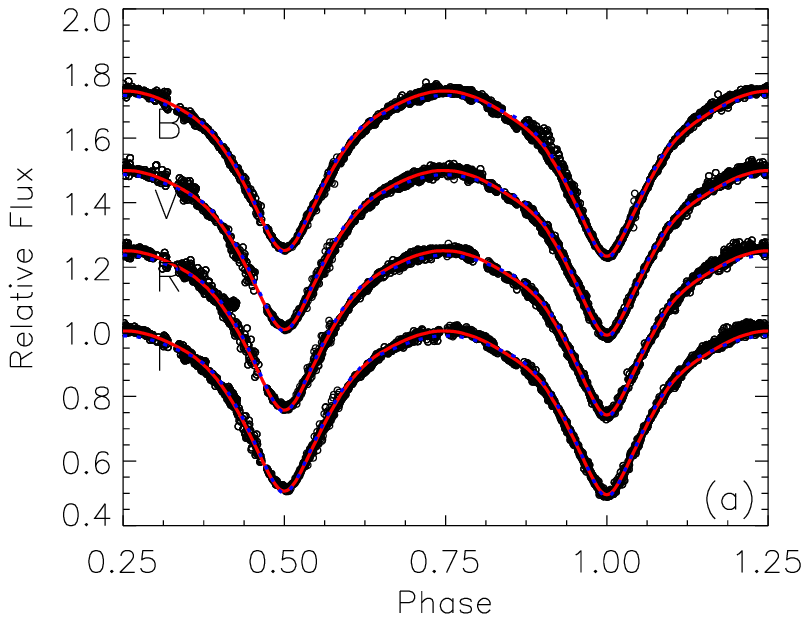}\\
\includegraphics[width=80mm]{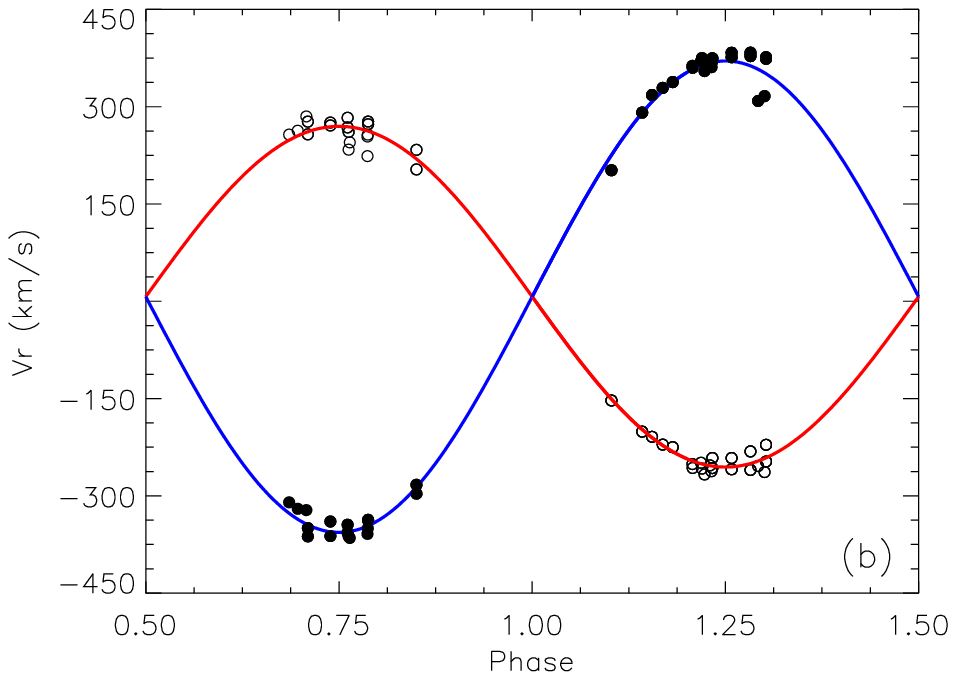}
\caption{(a) The observed and the computed light and (b) radial velocity curves of V382 Cyg.
The light curves in B, V, and R bands are altered by values of
0.75, 0.50, and 0.25, respectively, for a good visibility. See text for details.}\label{Fig:V382Cyg:LC}
\end{figure}

The observations were carried out on 25 nights in 2007-2011 with the 40-cm telescope that is equipped with the $2048\times2048$ Alta CCD camera at Ege University Observatory (EUO).
The exposure times are 15 s for B, 5 s for V, 4 s for R and 4 s for I filter.
The reduction and analysis of the frames have been
managed using the IRAF packages to subtract the bias and
dark then divide flat field, followed by aperture photometry
(DIGIPHOT/APPHOT).
Comparison and check stars are selected as HD 193204 (V=$8\fm32$, B-V=$0\fm46$), GSC 2684 1450 (V=$9\fm60$, B-V=$1\fm12$),
HD 228802 (V=$9\fm63$, B-V=$0\fm44$) and HD 193344 (V=$7\fm60$, B-V=$-0\fm06$).
The magnitudes and colors are obtained  from TYCHO-2 catalogue (H{\o}g et al. 2000).
We have studied all the nights and each frames separately during the data
reduction and standard deviations  are estimated for B, V, R and I bands as $0.018$, $0.013$, $0.016$ and $0.015$, respectively.
3095, 2861, 2871 and 2772 points are obtained in $B$, $V$, $R$ and $I$ bands, respectively.
We list all the observed data in Table~\ref{Tab:V382Cyg:LC}.
The light curves of the system obtained in this study are shown in Fig.~\ref{Fig:V382Cyg:LC}a.

We obtained three new minima times throughout these new observations.
They are collated with those published and listed in Table~\ref{tab:V382Cyg:mintimes} with their errors.
Using Table~\ref{tab:V382Cyg:mintimes} we derived a new linear ephemeris Eq.~(\ref{eq:v382cyg:1})
that we believe to be useful for future studies on the system:

\begin{equation}
\textrm{HJD\,MinI} = 24\,54344.4972(8)+1.8855270(2)\times E. \label{eq:v382cyg:1}
\end{equation}

\begin{table}
\caption{Observational points of V382 Cyg. Heliocentric Julian Date, phase, ($\Delta$m) and their corresponding filters are listed.
($\Delta$m) is the magnitude difference between variable (V382 Cyg) and comparison star (HD 193204). The phases were calculated using the Eq. (1). Table~\ref{Tab:V382Cyg:LC}  is given in its entirety in the electronic edition of
 this paper. A portion of it is shown here for guidance regarding its form and content.} \label{Tab:V382Cyg:LC}
\begin{tabular}{llll}
\hline
\hline
HJD	&	Phase	&	$\Delta$m	&	Filter \\	
\hline
2454292.33107&	0.3334&	0.331&	B\\
2454292.33444&	0.3352&	0.330&	B\\
2454292.33540&	0.3357&	0.324&	B\\
2454292.33636&	0.3362&	0.326&	B\\
2454292.33731&	0.3367&	0.325&	B\\
2454292.33826&	0.3372&	0.328&	B\\
2454292.33922&	0.3377&	0.315&	B\\
2454292.34018&	0.3382&	0.326&	B\\
2454292.34113&	0.3387&	0.332&	B\\
2454292.34209&	0.3392&	0.337&	B\\
2454292.34304&	0.3397&	0.330&	B\\
2454292.34400&	0.3403&	0.331&	B\\
2454292.34495&	0.3408&	0.338&	B\\
2454292.34592&	0.3413&	0.329&	B\\
2454292.34688&	0.3418&	0.327&	B\\
\hline
\end{tabular}
\end{table}

\section{Eclipse timings and period study}
The difference between the observed (O) and calculated (C) times of minima in an eclipsing binary system can provide us information about any orbital period variation.
The orbital period study of V382 Cyg was provided by Mayer et al. (1986), Mayer et al. (1998), De{\v g}irmenci et al. (1999) and Qian et al. (2007).
Using times of minima  Mayer et al. (1986) deduced the linear elements of the system.
De{\v g}irmenci et al. (1999) taking into account only the mass transfer between the components derived the mass
transfer rate from the less massive component to the more massive one as $5.0\times 10 ^{-6}$ M$_{\odot}$yr$^{-1}$.
Qian et al. (2007) analyzed the O-C diagram assuming both mass transfer ($\dot{M}=4.3\times10^{-7}$ M$_\odot$yr$^{-1}$) and a third body.

\begin{figure}
\includegraphics[width=80mm]{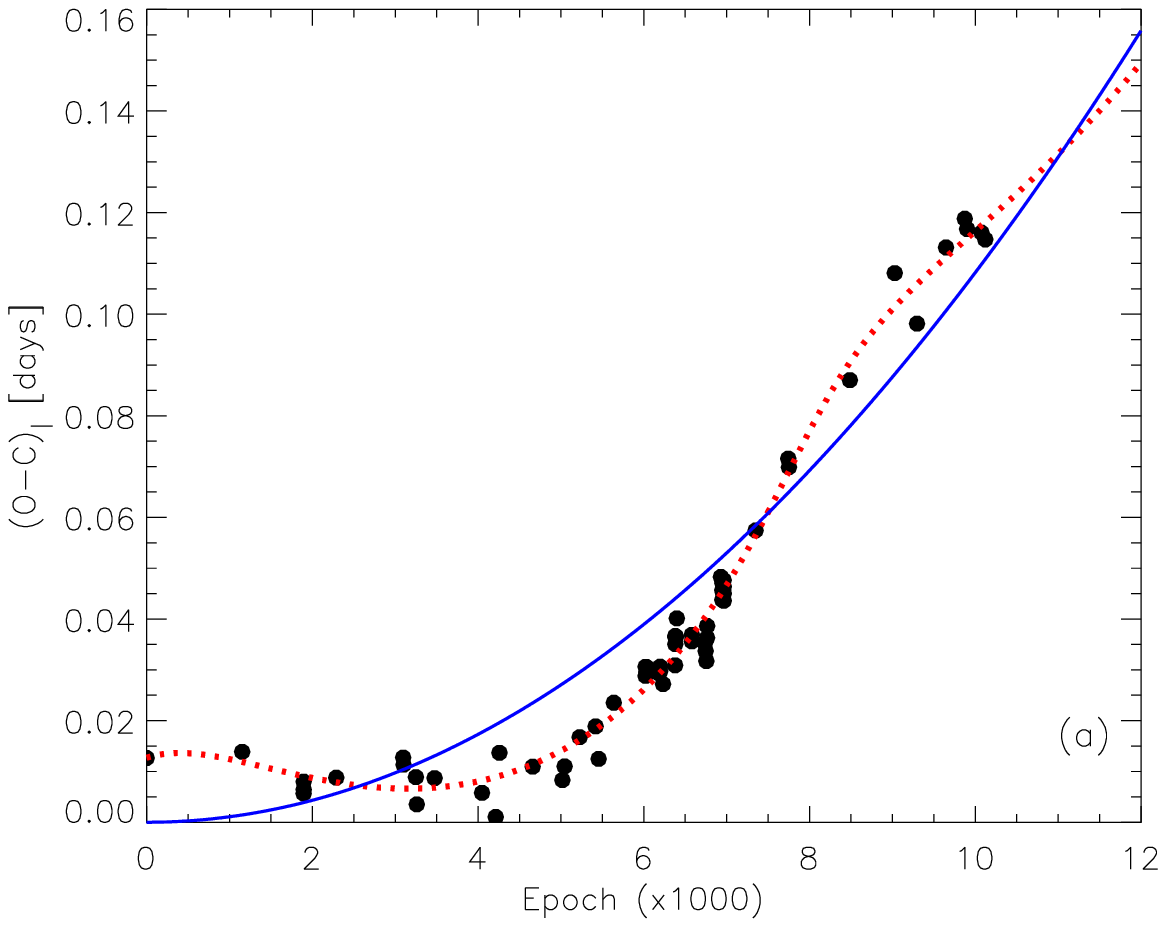} \\
\includegraphics[width=80mm]{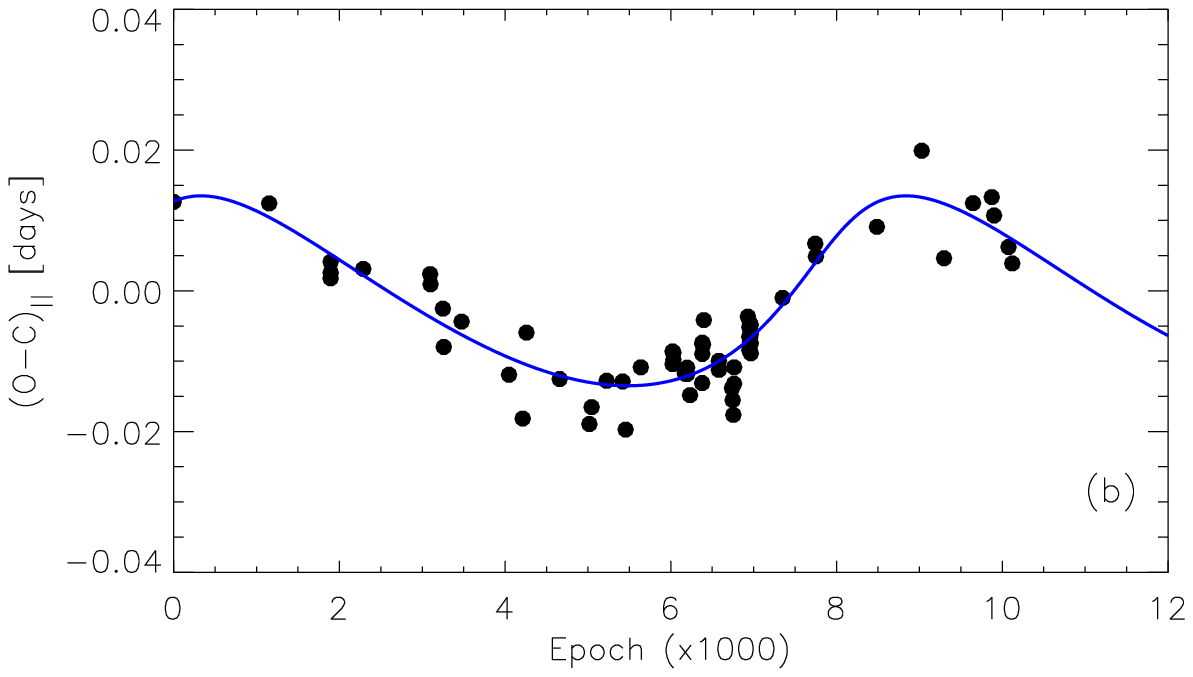} \\
\includegraphics[width=80mm]{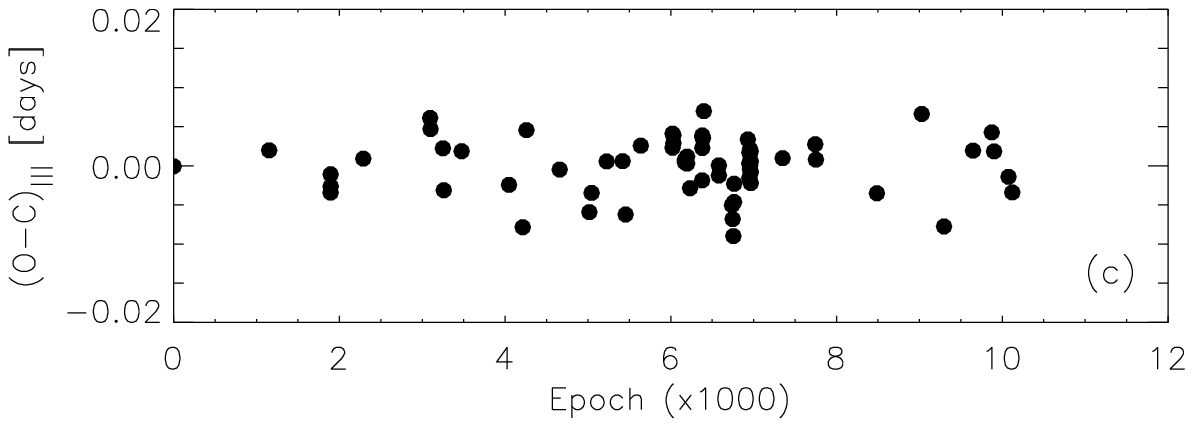}
\caption{ (a) Residuals for the times of minimum
light of V382 Cyg. The solid line is obtained with the quadratic
terms and the dotted line is obtained with parabolic+sinus terms in ephemeris Eq.~1. (b) The difference between the
observations and the quadratic ephemeris and (c) final residuals.}\label{Fig:V382Cyg:OC}
\end{figure}

Unlike previous studies of V382 Cyg, in this study we used only the most accurate (photoelectric and CCD data) minima times.
We solve the O-C curve to find a parabolic variation which can be ascribed as mass transfer
from the less massive companion to the more massive one. 
A total of 65 data points obtained with photometric/CCD observations are
used to study the period variation of the system.
Table~\ref{tab:V382Cyg:mintimes} shows the data used during the analysis.
The weighted least-squares method is used in order to determine the parameters of the upward
parabolic variation. The residuals (O-C)$_\textrm{I}$ indicate a quadratic solution (Fig.~\ref{Fig:V382Cyg:OC}a).
In order to estimate the light elements, the differential correction method is used.
The residuals (O-C)$_\textrm{II}$ show a sinusoidal variation (Fig.~\ref{Fig:V382Cyg:OC}b).
Therefore, we assumed a sinusoidal variation superposed on an upward parabola during O-C analysis.
A sine-like variation in the O-C curve, where both the primary and the
secondary minima follow the same trend, suggests the light time
effect via the presence of a tertiary component. We used Eq.~(\ref{Eq:V382cyg:1}) during the O-C analysis.
 \begin{equation}
\begin{array}{l}
\textrm{MinI} = T_o +P_oE + \frac{1}{2} \frac{dP}{dE}E^2  \\
\\
+ \frac {a_{12} \sin
i'}{c}  \left[ \frac {1-{e'}^2}{1+{e'} \cos v'} \sin \left( v' +
\omega' \right) +{e'} \sin \omega' \right]
 \label{Eq:V382cyg:1}
\end{array}
\end{equation}
where $T_{\rm o}$ is the starting epoch for the primary minimum, $E$ is the integer
eclipse cycle number, $P_{\rm o}$ is the orbital period of the
eclipsing binary and $a_{12}$, $i'$, $e'$ and $\omega'$
are the semi-major axis, inclination, eccentricity and the
longitude of the periastron of eclipsing pair about the third body
and $v'$ denotes the true anomaly of the position of the center of
mass. Time of periastron passage $T'$ and orbital period $P'$ are
the unknown parameters in Eq.(\ref{Eq:V382cyg:1}) (\emph{see} Kalomeni et al. 2007 for details).

\begin{table}
\begin{center}
\scriptsize
\caption{Times of minimum light for V382 Cyg.}\label{tab:V382Cyg:mintimes}
\begin{tabular}{llllll}
\hline
HJD* Min   &   Ref &   HJD Min    &   Ref &   HJD Min    &   Ref\\
\hline
2436814.7725 &	1 & 2448167.4765 & 7   &	2449913.4766 &	17\\
2438987.8298 &	1 &	2448186.3306 &	7  &	2449913.4785 &	10\\
2440385.9310 &	1 &	2448186.3316 &	7  &	2449915.3654 &	17\\
2440386.8761 &	1 &	2448447.4743 &	8  &	2449929.5074 &	17\\
2440387.8173 &	1 &	2448447.4764 &	8  &	2449930.4476 &	17\\
2441129.7698 &	1 &	2448498.3835 &	8  &	2449931.3909 &	17\\
2442651.3844 &	2 &	2448498.3844 &	8  &	2449945.5342 &	17\\
2442659.8678 &	3 &	2448564.3740 &	9  &	2449946.4757 &	17\\
2442940.8071 &	4 &	2448839.6629 &	10 &	2449947.4157 &	18\\
2442961.5424 &	3 &	2448843.4381 &	11 &	2449947.4171 &	17\\
2443366.9333 &	5 &	2448843.4395 &	12 &	2450671.4673 &	19\\
2444445.4450 &	6 &	2448843.4397 &	12 &	2451413.4316 &	20\\
2444757.4930 &	6 &	2448859.4664 &	11 &	2451429.4568 &	20\\
2444842.3538 &	6 &	2448878.3252 &	10 &	2452817.2130 &	21\\
2445598.4426 &	5 &	2449221.4844 &	13 &	2453837.2977 &	22\\
2446274.3971 &	5 &	2449221.4857 &	13 &	2455430.5686 &	23\\
2446325.3087 &	5 &	2449518.4527 &	14 &	2455483.361  &	24\\
2446668.4782 &	5 &	2449534.4780 &	15 &	2455811.4399 &	25\\
2447030.4922 &	3 &	2449550.5029 &	14 &	2454344.4913(4)&26\\
2447099.3141 &	3 &	2449567.4771 &	16 &	2455003.4938(5) &26\\
2447444.3744 &	3 &	2449568.4222 &	16 &	2455897.2295(3)&26\\
2448167.4747 &	7 & 2449880.4846 &	17 &				   & \\	
\hline
\end{tabular}
\end{center}
\scriptsize {References for Table~\ref{tab:V382Cyg:mintimes}.
1-  Landolt(1975)
2-  Mayer(1980)
3-  Mayer et al.(1991)
4-  Bloomer, Burke, \& Millis (1979)
5-  Mayer et al.(1986)
6-  Andrakakou et al.(1980)
7-  H\"ubscher, Agerer, \& Wunder(1991)
8-  H\"ubscher, Agerer, \& Wunder(1992)
9-  Bl\"attler(1992)
10- Mayer et al.(1998)
11- Peter(1992)
12- H\"ubscher, Agerer, \& Wunder(1993)
13- H\"ubscher, Agerer, \& Wunder(1994)
14- Peter (1994)
15- Agerer \&  H\"ubscher(1995)
16- Blaettler(1994)
17- De\v girmenci et al.(1999)
18- Agerer \& H\"ubscher(1996)
19- Agerer \&  H\"ubscher(1998)
20- Agerer \&  H\"ubscher(2001)
21- Nagai(2004)
22- Qian et al.(2007)
23- H\"ubscher(2011)
24- Paschke(2010)
25- Zasche et al.(2011)
26- present study.}
\end{table}

\begin{table*}
\caption{Orbital elements of the tertiary component and mass transfer ratio for V382 Cyg. The
standard errors 1$\sigma$, in the last digit are given in
parentheses.} \label{tab:V382:OCResults}
\begin{tabular}{lll}
\hline
Parameter            &Unit               & Value                       \\
\hline $T_o$         & [HJD]             & 2436814.7599(59)             \\
$P_o$                & [day]             & 1.885515(2)                 \\
$P'$                 & [year]            & 43.9(1.7)                        \\
$T'$                 & [HJD]             & 2419843(1611)                      \\
$e'$                 &                   & 0.41(8)                          \\
$\omega'$            & [$^\circ$]        & 34(15)                              \\
$a_{12} \sin i'$     & [AU]              & 2.49(14)                               \\
$f(m)$               & [M$_{\odot}$]     & 0.0080(2)                                  \\
$m_{3;i'=30^\circ}$  & [M$_{\odot}$]     & 5.7                                        \\
$m_{3;i'=90^\circ}$  & [M$_{\odot}$]     & 2.8                                      \\
$\frac{1}{2}\frac{dP}{dE}$ & [c/d]       & $1.08(1)\times 10^{-9}$              \\
\hline
\end{tabular}
\end{table*}

\section{Simultaneous solutions of light and radial velocity curves}

The $\textsc{BVRI}$ light curves obtained in this study are solved simultaneously with the radial velocity curve of Harries et al. (1997) and Mayer et al. (2002)
using {\sc Phoebe} (Pr\~{s}a \& Zwitter 2005) which is based on the WD code (Wilson \& Devinney 1971; Wilson 1979; Wilson 1990).
During the solution we used weighted light curves that are constructed according to the standard deviation of each filter.
The effective temperature of the hot star is chosen according to spectroscopic studies.
The albedos values $A_{\rm 1}$ and $A_{\rm 2}$, for the hot and cooler components are adopted from Rucinski (1969),
whilst the values of the gravity darkening coefficients $g_{\rm 1}$ and $g_{\rm 2}$ are taken from von Zeipel (1924).
The logarithmic limb-darkening law is used with the coefficients adopted from van Hamme (1993)
 for a solar composition star. The adjustable photometric parameters are
 orbital inclination, \textit{i}, surface potential, $\Omega_{1,2}$, temperature
 of the secondary component $T_{\rm 2}$, luminosity $L_{\rm 1}$ and the mass ratio $q$. The center of
 mass velocity $V_0$ and semi-major axis $a$ are also set as free parameters as well as the time of minimum light,
 $T_0$ and the orbital period, $P$. Table~\ref{tab:V382Cyg:lcanalysis} summarizes the result of the analysis. $B$, $V$, $R$ and $I$
 light curves and the velocity curve computed using the determined  parameters are shown by solid lines in
 Fig.~\ref{Fig:V382Cyg:LC}a.

During solution we assumed both contact (C) and semi-detached (SD) configurations.
We emphasized that both solutions have similar results (see Table~\ref{tab:V382Cyg:lcanalysis}).
In Fig.~\ref{Fig:V382Cyg:LC} the light curve fits for a contact and a semi-detached configurations are shown with a dotted and a solid line, respectively.
For the contact model the filling factor, which is expressed by $(\Omega _{\textrm{in}} -\Omega)/ (\Omega _{\textrm{in}} -\Omega _{\textrm{out}} )$
that varies from zero to unity from the inner (${\Omega _{\textrm{in}}}$) to the outer critical surface ($\Omega _{\textrm{out}}$), is calculated to be $f = 0.09$.
New solutions indicate that the system has a weak contact
degree or that it is a semi-detached binary with a near-contact
configuration.

\begin{table}
\scriptsize
\caption{Simultaneous analyses results of the light and radial velocity curve and their formal 1$\sigma$ errors for V382~Cyg.
The column headers C and SD refer to the contact model and semi-detached model, respectively. See text for details.}
\begin{tabular}{lll}
\hline
Parameter                                   & C              & SD                  \\
\hline
$i$ ${({^\circ})}$                          & 84.5(1)        & 85.3(2)            \\
$q = M_h / M_c$                             & 0.7439(14)     & 0.745(2)          \\
$a$ ($\rm{R_{\odot}}$)			            & 23.45(13)      & 23.47(12)          \\
$V_0$ (kms$^{-1}$)			                & 7.1(1.6)       & 6.3(1.6)           \\
$\Omega _{1}$                               & 3.275(3)       & 3.390(9)           \\
$\Omega _{2}$                               & 3.275(3)       & 3.320             \\
$T_1$ (K)                                   & 36000          & 36000              \\
$T_2$ (K)                                   & 34415(270)     & 34578(240)         \\
Fractional radius of primary comp.          & 0.4139(5)      & 0.4087(2)          \\
Fractional radius of secondary comp.        & 0.3604(6)      & 0.3526(2)          \\
$A_1 = A_2$                                 & 1.0            & 1.0                \\
$g_1 = g_2$                                 & 1.0            & 1.0                \\
Luminosity ratio:$\frac{L_1}{L_1+L_2}$ (\%)     &            &               \\
$B$                                         & 57.9           & 56.2              \\
$V$                                         & 58.0           & 56.3              \\
$R$   					                    & 57.7           & 56.1              \\
$I$   					                    & 57.5           & 55.8               \\
\hline
\end{tabular}
\label{tab:V382Cyg:lcanalysis}
\end{table}


\section{Results and Conclusion}

Newly obtained B, V, R and I light curves of the early-type interacting system V382 Cyg have been solved with earlier published spectroscopic studies.
We derived the orbital and the physical parameters of the components  from simultaneous solution and listed the physical parameters in Table~\ref{tab:V382Cyg:par}.
We calculated the masses as 27.9 $\rm{M_{\odot}}$  and 20.8 $\rm{M_{\odot}}$ which are $\sim7\%$ higher than those given by
Harries et al. (1997) and  $\sim7\%$ lower than Mayer et al. (2002) results.
The distance of the system to the Sun is estimated as $1455\mp76$ pc using the observed parameters.
During the calculations bolometric corrections are taken from Lanz \& Hubeny (2003),
the effective temperature and absolute magnitude of the Sun are taken as 5777~K and 4.732 mag.
Garmany \& Stencel (1992) proposed that V382 Cyg may be a member of the Cyg OB1 association.
The distance to Cyg OB1 reported by Uyan{\i}ker et al. (2001) in the range from 1.25 kpc to 1.83 kpc.
In this study, the distance of V382 Cyg was found to be within this range, so it is likely to be a member of Cyg OB1.

We have collected 65 times of mid-eclipses including our three new ones to search any periodic variation.
Analysis of O--C residuals can give us the reason of a periodic and/or non-periodic variation.
We applied the least-square method to the Eq.~(\ref{Eq:V382cyg:1}) to find the period change rate.
Then the mass transfer rate is calculated and the possible third body parameters are estimated (Table~\ref{tab:V382:OCResults}).
According our analysis, there is a mass transfer from the less massive (initially massive one) to the more massive
component ($\frac{dM}{dt}=6.1\times 10^{-6} \rm{M_{\odot}}/yr$) and the period change rate ($\frac{P}{\dot{P}}=4.5\times 10^6$).
In this study, the obtained mass transfer ratio is nearly ten times higher than that of Qian et al. (2007). This difference is probably due to their formalism.
Additionally, we found new parameters for the tertiary component with 43.9 yrs outer period.

Isolated massive stars and massive stars in  binary  systems (e.g. V382 Cyg) evolve in different ways.
The evolutionary path towards the final supernovae will become significantly different.
The nature of the final stellar remnant may also be altered so that while a neutron star might be expected,
mass transfer may lead to a black hole formation.
Observations of systems similar to V382 Cyg can allow us to reveal the nature of the binary evolution
and may help us to understand the possible variation of stellar lifecycles (see Eggleton 2010, Eldridge \& Stanway 2009 and Yakut et al. 2013).
Using our newly obtained physical parameters we provide an evolution model of the massive interacting binary system V382 Cyg.
We used the TWIN version of the EV code (Yakut \& Eggleton 2005, Eggleton 2008, Eggleton 2010) that has been  developed by Peter P. Eggleton.

\begin{table*}
\begin{center}
\caption{Astrophysical parameters of the system. The standard errors
1$\sigma$ in the last digit are given in parentheses.}
\label{tab:V382Cyg:par}
\begin{tabular}{llll}
\hline
Parameter                                        &Unit                      & Primary           & Secondary   \\
\hline
Mass (M)                                         &$\rm{M_{\odot}}$        & $27.9(5)$            & $20.8(4)$      \\
Radius (R)                                       &$\rm{R_{\odot}}$        & $9.7(2)$             & $8.5(2)$      \\
Temperature (T$_{\rm eff}$)                      &$\rm{K}$                & 36000                & 34415(270)     \\
Luminosity (L)                                   &$\log{(L_*/L_{\odot})}$ & $5.152(20)$          & $4.954(19)$      \\
Surface gravity ($\log g$)                       &$\rm{cms^{-2}} $        & $3.91$               & $3.90$      \\
Bolometric magnitude (M$_b$)                     &mag                     & -8.15                & -7.65           \\
Absolute magnitude (M$_V$)                       &mag                     & -4.75                & -4.38           \\
Period change rate ($\dot{P}$)                   &d/yr                   &~~~~~~~~~~~~~$4.2(3)\times10^{-7}$ &      \\
Mass transfer ratio ($\dot{M}$)                  &M$_\odot$/yr           &~~~~~~~~~~~~~$6.1(4)\times10^{-6}$ &      \\
Distance (d)                                     &pc                      &~~~~~~~~~~~~~~~~~1466(76) &      \\
\hline
\end{tabular}
\end{center}
\end{table*}

\begin{figure}
\includegraphics[width=80mm]{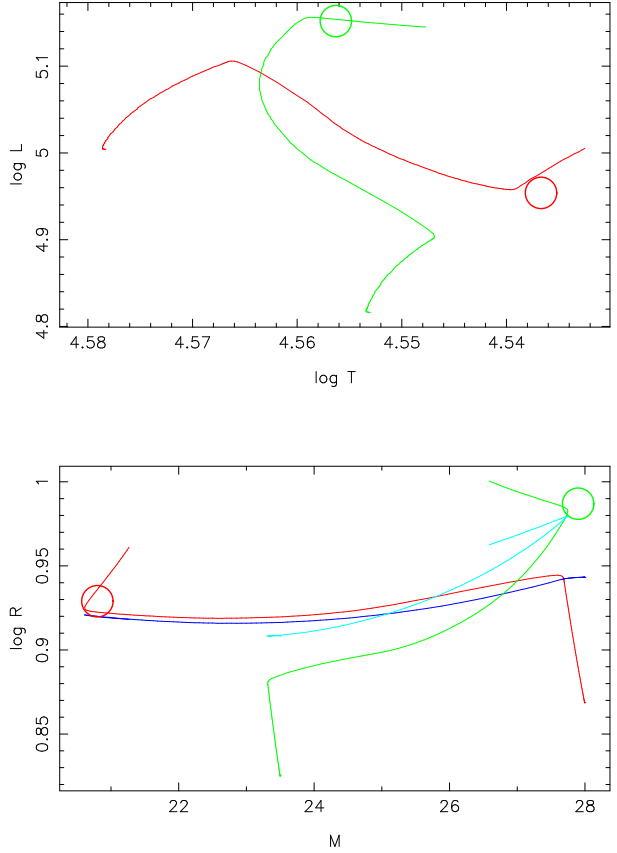}
\caption{Non-conservative evolution of V382 Cyg in the $H-R$ diagram (top panel) and $\log R\ vs.\
M$ plane (button panel). More massive and less massive star are shown in red and green; their respective
Roche-lobe radii are dark blue and light blue.
Initial parameters are 28.0 M$_{\odot}$, 23.5 M$_{\odot}$ and 1.72 days with assumed Solar composition.
The original primary is the secondary currently.}\label{fig:model}
\end{figure}

We run a few models using different initial parameters.
Among these models the best agreement with the observations is obtained for the model binary system whose initial
period is 1.72 days, with a primary and secondary masses  28 M$_\odot$ and 23.5 M$_\odot$.
The first 28 models shrink slightly since the ZAMS is somewhat artificial.
The massive component fills its Roche lobe and RLOF begins at model 123.
At model 201 the parameters are very consistent with the current observed parameters (Table~\ref{tab:V382Cyg:model}).
After two more models it reaches a contact phase. This configuration is consistent with the O-C analysis (\textit{see } \S 3).
Therefore, our models show that the system looks like a semi-detached system but very close to a contact phase.
We plotted our result in the H-R diagram and Mass-Radius planes (Fig.~\ref{fig:model}).
We found the age of the system to be 3.85 Myr which is in the range of age of Cygnus OB1 (e.g., Massey et al. 1995).

\begin{table*}
\begin{center}
\caption{Observed and model parameters of the components.}
\label{tab:V382Cyg:model}
\begin{tabular}{llllllll}
\hline
             & Star &M (M$_\odot$) &  log R  & log T   &  log L  &     P (days)   & Age  (yr) \\
\hline
Observed     & A &   20.8       &  0.9269 &  4.5367 &  4.954  &   1.8855       &       \\
             & B &   27.9       &  0.9868 &  4.5563 &  5.152  &                &       \\
\hline
Model No 28  & A &   28.0       &  0.8687 &  4.5783 &  5.0039 &   1.6960       & 8.6$\times 10^3$  \\
             & B &   23.5       &  0.8252 &  4.5532 &  4.8163 &                &       \\
\hline
Model No 201 & A &   20.83      &  0.9224 &  4.5401 &  4.9581 &   1.8860       & 3.85$\times 10^6$    \\
             & B &   27.65      &  0.9741 &  4.5606 &  5.1440 &                &         \\
\hline
\end{tabular}
\end{center}
\end{table*}

\section*{Acknowledgments}
The authors would like to thank Peter P. Eggleton for his valuable comments and suggestions to improve the quality of the paper and
for providing the current version of the EV(TWIN) code and, B. Kalomeni for helpful comments.
We are very grateful to an anonymous referee for his/her comments and helpful constructive suggestions which helped us to improve the paper.
This study was supported by the Turkish Scientific
and Research Council (T\"UB\.ITAK 111T270) and the Ege University Research Fund.
KY acknowledges support by the The Turkish Academy of Sciences (T\"UBA).
The current study is a part of PhD thesis of B. Ya{\c s}arsoy.


\begin{thebibliography}{}
\bibitem[\protect\citeauthoryear{Agerer \& Hubscher}{1995}]{1995IBVS.4222....1A} Agerer, F. \& H\"ubscher, J. 1995, IBVS, 4222, 1
\bibitem[\protect\citeauthoryear{Agerer \& Hubscher}{1996}]{1996IBVS.4383....1A} Agerer, F. \& H\"ubscher, J. 1996, IBVS, 4383, 1
\bibitem[\protect\citeauthoryear{Agerer \& Huebscher}{1998}]{1998IBVS.4606....1A} Agerer, F. \& H\"uebscher, J. 1998, IBVS, 4606, 1
\bibitem[\protect\citeauthoryear{Agerer \& Hubscher}{2001}]{2001IBVS.5016....1A} Agerer, F. \& H\"ubscher, J. 2001, IBVS, 5016, 1
\bibitem[\protect\citeauthoryear{Andrakakou et al.}{1980}]{1980BBSAG..49....1A} Andrakakou, M., et al. 1980, BBSAG, 49, 1
\bibitem[\protect\citeauthoryear{Blaettler}{1994}]{1994BBSAG.107....1B} Blaettler, V. 1994, BBSAG, 107, 1
\bibitem[\protect\citeauthoryear{Bl{\"a}ttler}{1992}]{1992BBSAG.99...1B} Bl{\"a}ttler, E. 1992, BBSAG, 99, 1
\bibitem[\protect\citeauthoryear{Bloomer, Burke, \& Millis}{1979}]{1979BAAS...11..439B} Bloomer, R.~H., Burke, E.~W. \& Millis, R.~L. 1979, BAAS, 11, 439
\bibitem[\protect\citeauthoryear{Burkholder, Massey, \& Morrell}{1997}]{1997ApJ...490..328B} Burkholder, V., Massey, P. \& Morrell, N. 1997, ApJ, 490, 328
\bibitem[\protect\citeauthoryear{Cester et al.}{1978}]{1978A&AS...33...91C} Cester, B., Fedel, B., Giuricin, G., Mardirossian, F. \& Mezzetti, M. 1978, A\&AS, 33, 91
\bibitem[\protect\citeauthoryear{De{\v g}irmenci et al.}{1999}]{1999A&AS..134..327D} De{\v g}irmenci, {\"O}.~L., Sezer, C., Demircan, O., Erdem, A., {\"O}zdemir, S., Ak, H. \& Albayrak, B. 1999, A\&AS, 134, 327
\bibitem[\protect\citeauthoryear{Eggleton}{2008}]{2008IAUS..246..228E} Eggleton, P.~P. 2008, IAUS, 246, 228
\bibitem[\protect\citeauthoryear{Eggleton}{2010}]{2010NewAR..54...45E} Eggleton, P.~P. 2010, NewAR, 54, 45
\bibitem[\protect\citeauthoryear{Eldridge}{2009}]{2009MNRAS.400L..20E} Eldridge, J.~J. 2009, MNRAS, 400, L20
\bibitem[\protect\citeauthoryear{Garmany \& Stencel}{1992}]{1992A&AS...94..211G} Garmany, C.~D. \& Stencel, R.~E. 1992, A\&AS, 94, 211
\bibitem[\protect\citeauthoryear{Harries, Hilditch, \& Hill}{1997}]{1997MNRAS.285..277H} Harries, T.~J., Hilditch, R.~W. \& Hill, G. 1997, MNRAS, 285, 277
\bibitem[H{\o}g et al.(2000)]{2000A&A...355L..27H} H{\o}g, E., Fabricius, C., Makarov, V.~V., et al.\ 2000, \aap, 355, L27
\bibitem[\protect\citeauthoryear{H{\"u}bscher, Agerer, \& Wunder}{1991}]{1991BAVSM..59....1H} H{\"u}bscher, J., Agerer, F., \& Wunder, E. 1991, BAVSM, 59, 1
\bibitem[\protect\citeauthoryear{H{\"u}bscher, Agerer, \& Wunder}{1992}]{1992BAVSM..60....1H} H{\"u}bscher, J., Agerer, F., \& Wunder, E. 1992, BAVSM, 60, 1
\bibitem[\protect\citeauthoryear{H{\"u}bscher, Agerer, \& Wunder}{1993}]{1993BAVSM..62....1H} H{\"u}bscher, J., Agerer, F., \& Wunder, E. 1993, BAVSM, 62, 1
\bibitem[\protect\citeauthoryear{H{\"u}bscher, Agerer, \& Wunder}{1994}]{1994BAVSM..68....1H} H{\"u}bscher, J., Agerer, F., \& Wunder, E. 1994, BAVSM, 68, 1
\bibitem[\protect\citeauthoryear{Hubscher}{2011}]{2011IBVS.5984....1H} H\"ubscher, J. 2011, IBVS, 5984, 1
\bibitem[\protect\citeauthoryear{Kalomeni et al.}{2007}]{2007AJ....134..642K} Kalomeni, B., Yakut, K., Keskin, V., De{\u g}irmenci, {\"O}.~L., Ula{\c s}, B. \& K{\"o}se, O.\ 2007, AJ, 134, 642
\bibitem[\protect\citeauthoryear{Koch, Siah,\& Fanelli}{1979}]{1979PASP...91..474K} Koch, R.~H., Siah, M.~J. \& Fanelli, M.~N. 1979, PASP, 91, 474
\bibitem[\protect\citeauthoryear{Landolt}{1964}]{1964ApJ...140.1494L} Landolt, A.~U. 1964, ApJ, 140, 1494
\bibitem[\protect\citeauthoryear{Landolt}{1975}]{1975PASP...87..409L} Landolt, A.~U. 1975, PASP, 87, 409
\bibitem[\protect\citeauthoryear{Lanz \& Hubeny}{2003}]{2003ApJS..146..417L} Lanz, T. \& Hubeny, I., 2003, ApJS, 146, 417
\bibitem[\protect\citeauthoryear{Massey, Johnson, \& Degioia-Eastwood}{1995}]{1995ApJ...454..151M} Massey, P., Johnson, K.~E. \& Degioia-Eastwood, K. 1995, ApJ, 454, 151
\bibitem[\protect\citeauthoryear{Mayer}{1980}]{1980BAICz..31..292M} Mayer, P. 1980, BAICz, 31, 292
\bibitem[\protect\citeauthoryear{Mayer et al.}{1986}]{1986IBVS.2942....1M} Mayer, P., Wolf, M., Muminovic, M. \& Stupar, M. 1986, IBVS, 2942, 1
\bibitem[\protect\citeauthoryear{Mayer et al.}{1991}]{1991BAICz..42..230M} Mayer, P., Hadrava, P., Harmanec, P. \& Chochol, D. 1991, BAICz, 42, 230
\bibitem[\protect\citeauthoryear{Mayer et al.}{1998}]{1998A&AS..130..311M} Mayer, P., Niarchos, P.~G., Lorenz, R., Wolf, M. \& Christie, G. 1998, A\&AS, 130, 311
\bibitem[\protect\citeauthoryear{Mayer, Lorenz, \& Drechsel}{2002}]{2002A&A...388..268M} Mayer, P., Lorenz, R. \& Drechsel, H. 2002, A\&A, 388, 268
\bibitem[\protect\citeauthoryear{Morgenroth}{1935}]{1935AN....255..425M} Morgenroth, O. 1935, AN, 255, 425
\bibitem[\protect\citeauthoryear{Nagai}{2004}]{2004Var.Star Bull.Japan..42....1N} Nagai, K. 2004, Var.Star Bull.Japan, 42, 1
\bibitem[\protect\citeauthoryear{Paschke}{2010}]{2010OEJV..130....1P} Paschke, A. 2010, OEJV, 130, 1
\bibitem[\protect\citeauthoryear{Pearce}{1952}]{1952PASP...64..219P} Pearce, J.~A. 1952, PASP, 64, 219
\bibitem[\protect\citeauthoryear{Peter1}{1992}]{1992BBSAG..102....1A} Peter, H. 1992, BBSAG Bull., 102
\bibitem[\protect\citeauthoryear{Peter2}{1994}]{1994BBSAG..107....1A} Peter, H. 1994, BBSAG Bull., 107
\bibitem[\protect\citeauthoryear{Petrov}{1946}]{1946PeremZvezdy6..72P} Petrov, A.A. 1946, Perem. Zvezdy, 6, 72
\bibitem[\protect\citeauthoryear{Popper}{1978}]{1978ApJ...220L..11P} Popper, D.~M. 1978, ApJ, 220, L11
\bibitem[\protect\citeauthoryear{Popper}{1980}]{1980ARA&A..18..115P} Popper, D.~M. 1980, ARA\&A, 18, 115
\bibitem[\protect\citeauthoryear{Popper \& Hill}{1991}]{1991AJ....101..600P} Popper, D.~M., \& Hill G. 1991, AJ, 101, 600
\bibitem[Pr{\v s}a \& Zwitter(2005)]{2005ApJ...628..426P} Pr{\v s}a, A., \& Zwitter, T.\ 2005, ApJ, 628, 426
\bibitem[\protect\citeauthoryear{Qian et al.}{2007}]{2007MNRAS.380.1599Q} Qian, S.-B., Yuan, J.-Z., Liu, L., He, J.-J. \& Fern{\'a}ndez Laj{\'u}s, E., \&
Kreiner, J.~Z. 2007, MNRAS, 380, 1599
\bibitem[Rucinski(1969)]{rucinski69} Rucinski, S. M. 1969, Acta Astron. 19, 245
\bibitem[\protect\citeauthoryear{Uyan{\i}ker et al.}{2001}]{2001A&A...371..675U} Uyan{\i}ker, B., F{\"u}rst, E., Reich, W., Aschenbach, B., \& Wielebinski, R. 2001, A\&A, 371, 675
\bibitem[van Hamme(1993)]{1993AJ....106.2096V} van Hamme, W.\ 1993, AJ, 106, 2096
\bibitem[\protect\citeauthoryear{von Zeipel}{1924}]{ze}von Zeipel, H. 1924, MNRAS, 84, 665
\bibitem[\protect\citeauthoryear{Wilson and Devinney}{1971}]{wi}Wilson, R.E., \& Devinney, E.J. 1971, ApJ, 166, 605
\bibitem[\protect\citeauthoryear{Wilson}{1979}]{wi2}Wilson, R.E. 1979, ApJ, 234, 1054
\bibitem[\protect\citeauthoryear{wilson}{1990}]{wi3}Wilson, R.E. 1990, ApJ, 356, 613
\bibitem[Yakut \& Eggleton(2005)]{ky05} Yakut, K., \& Eggleton, P.~P.\ 2005, ApJ, 629, 1055
\bibitem[\protect\citeauthoryear{Yakut et al.}{2013}]{2013} Yakut, K., et al., 2013, in preparation
\bibitem[\protect\citeauthoryear{Zasche et al.}{2011}]{2011IBVS.6007....1Z} Zasche, P., Uhlar, R., Kucakova, H., \& Svoboda, P. 2011, IBVS, 6007, 1
\end{thebibliography}
\end{document}